\documentclass[
preprint,
 amsmath,amssymb,
 aps,
pra,
]{revtex4-1}

\usepackage{graphicx}
\usepackage{dcolumn}
\usepackage{bm}

\begin{document}

\title{Predicting Atomic Decay Rates Using an Informational-Entropic Approach}

\author{Marcelo Gleiser}
\email{Marcelo.Gleiser@dartmouth.edu}

\author{Nan Jiang}
\email{Nan.Jiang.GR@dartmouth.edu}

 \affiliation{Department of Physics and Astronomy, Dartmouth College, Hanover, NH 03755, US}
\date{\today}

\begin{abstract}
We show that a newly proposed Shannon-like entropic measure of shape complexity applicable to spatially-localized or periodic mathematical functions known as configurational entropy (CE) can be used as a predictor of spontaneous decay rates for one-electron atoms. The CE is constructed from the Fourier transform of the atomic probability density. For the hydrogen atom with degenerate states labeled with the principal quantum number $n$, we obtain a scaling law relating the $n$-averaged decay rates to the respective CE. The scaling law allows us to predict the $n$-averaged decay rate without relying on the traditional computation of dipole matrix elements. We tested the predictive power of our approach up to $n=20$, obtaining an accuracy better than 3.7\% within our numerical precision, as compared to spontaneous decay tables listed in the literature.
\end{abstract}

\pacs{Valid PACS appear here}

\maketitle


\section{Introduction}

From subatomic to cosmological scales, investigating the stability of spatially-bound physical systems is a fundamental test of the efficacy of mathematical models describing natural phenomena. In general, the system's stability is tested against variations of one or more parameters controlling its overall physical properties. For example, by tuning the self-interaction energy in Bose-Einstein condensates \cite{cornish2000}, or by changing a star's central energy-density while keeping the number of baryons constant \cite{Shapiro}. In classical systems, perturbation techniques are used to find instability regions in parameter space, with growing (oscillating) modes indicating instability (stability) \cite{hayashi2014nonlinear}. In quantum systems the situation is different due to the possibility of spontaneous decay of excited states: atoms which are away from their ground states may decay due to impinging electromagnetic vacuum fluctuations. In this case, it could be argued that instability is not inherent to the system's internal dynamics, as in the classical case, but induced by its interactions with the surrounding environment \cite{berestetskii1982quantum}. However, as is well-known, the decay rate expression for spontaneous emission does not make reference to an external electromagnetic field (as is the case with stimulated emission, or with quantum decoherence for a variety of other ``environmental influences'' \cite{zurek2003decoherence}), depending only on the properties of the atomic eigenfunctions of the final and initial states. For the simplest case of one-electron atoms these are known, and the theory of spontaneous decay is one of the great successes of quantum physics.

In the present paper, we revisit spontaneous decay from a novel perspective, that of information theory. We use a recently-proposed extension of Shannon's information theory \cite{Shannon} applied to spatially-bound or periodic physical systems known as Configurational Entropy (CE) \cite{GS1} to estimate the lifetimes of excited states of one-electron atoms. ``Information'' here must be considered in the proper context, defined briefly in the next Section. More details can be found in Ref. \cite{GSo3}. The ``message'' is a particular configuration described by a spatially-bound or periodic function, in this case the eigenfunctions of the hydrogen atom Hamiltonian for the quantum numbers ${n,\ell,m}$. The ``alphabet'' is given by the momentum modes which compose the configuration with specific weights (probabilities), as obtained from its Fourier transform. This way, each eigenfunction -- or message -- will have a specific informational signature in momentum space with quantifiable complexity. 

Shannon's information entropy gives a measure of the complexity of a language in terms of its compressibility (or redundancy): the more redundant and thus compressible a language, the lower its information entropy. Messages in languages with high compressibility require less bits of information to encode. This is consistent with the interpretation of information entropy as a measure of our ignorance about the system. Thermodynamically, a system in equilibrium maximizes entropy because we have no information about its initial condition (except for conserved quantities): equipartition means no memory. Analogously, the configurational entropy measures the relative spatial complexity of the function describing the physical system in terms of its momentum-mode decomposition. A sinusoidal wave requiring a single momentum mode minimizes CE. In general, configurations with more spatial localization require more momentum modes with larger relative amplitudes and thus have higher configuration entropy. These statements will be made explicit below. Given that the level of spatial confinement of a given function depends on the details of the physics it describes (interactions, boundary conditions), the specific physics that defines the function's spatial properties is implicitly encoded in its CE.

Proposed in 2012, configurational entropy has been applied to several physical systems, including solitons in field theory \cite{GS1}, relativistic and nonrelativistic stars \cite{GSo1,GJ1}, and phase transitions \cite{GSo2,GSo3}, among other applications in high energy physics and cosmology \cite{correa2015entropic, gleiser2011generation}. This paper opens a new front, applying CE to atomic physics, in particular to unstable atomic states. In Section \ref{CEsec} we review the basic concepts needed to apply configuration entropy to atomic states. In Section \ref{CEHatom} we define and compute the CE for different atomic states of the hydrogen atom. Comparing the value of CE with the lifetime for $n$-averaged states, we obtain a simple scaling relation that allows us to use the CE as a predictor of the $n$-averaged atomic-state lifetimes with an accuracy smaller than 3.7\% for states with $n\leq 20$. Since we see no growing deviation from the scaling law with increasing $n$, we can extrapolate its validity to higher values of $n$. In Section \ref{Concluding} we summarize our results and suggest possible future applications, while in the Appendix we present technical details of the calculation and of our numerical approach.

\section{Configurational Entropy}
\label{CEsec}
\noindent We follow the definition of configuration entropy (CE) of Gleiser and Stamatopoulos \cite{GS1}, which we briefly repeat here for convenience.  Consider a continuous square-integrable function $g(\mathbf{x})$ defined on $\mathbf{R^d}$ with Fourier transform
\begin{equation}
\label{FK}
G(\mathbf{k}) = \int_{\mathbf{R^d}}\exp(-i\mathbf{k}\cdot\mathbf{x})g(\mathbf{x})d^d\mathbf{x}.
\end{equation}
\noindent
Now introduce the modal fraction,
\begin{equation}
\label{fk}
f(\mathbf{k}) = \frac{|G(\mathbf{k})|^2}{\int |G(\mathbf{k})|^2 d^d\mathbf{k}}.
\end{equation}
The configurational entropy is then defined as
\begin{equation}
    \label{CE}
    S_c[\tilde{f}(\mathbf{k})] = -\int_{\mathbf{R^d}} \tilde{f}(\mathbf{k})\log \tilde{f}(\mathbf{k}) d^d\mathbf{k},
\end{equation}
\noindent
where we normalized the modal fraction over the mode that carries maximum weight, $f_{\rm max}(\mathbf{k})$, as
\begin{equation}
\label{nfk}
\tilde{f}(\mathbf{k}) = f(\mathbf{k})/f_{\rm {max}}(\mathbf{k}).
\end{equation}
This normalization guarantees the positivity of the configurational entropy. The integrand in Eq. \ref{CE} is called the configuration entropy density (CE density). For periodic functions, one would use instead the Fourier series for the function $g(\mathbf{x})$ and define the discrete modal fraction as $f_n = |a_n|^2/\sum |a_n|^2$, where the $\{a_n\}$ are the relative weights for the different modes $n$. We note that it should be possible, in principle, to choose other functional transforms to obtain alternative definitions for the configurational entropy. We choose the Fourier transform due to its clear physical interpretation, as it relates increased spatial localization with a wider momentum-mode distribution.

As explained in Ref. \cite{GSo3}, there is a clear connection between CE and Shannon's information entropy, widely used in the context of message transmission and decoding \cite{Shannon}. Recall that in Shannon's formula, entropy is maximized when all symbols (of an alphabet) have the same average probability of appearing. This is also the state of maximal ignorance or uncertainty, with minimal correlation between adjacent symbols. 

As mentioned in the Introduction, in the case of CE we can {\it informally} interpret each mode of a field configuration as a ``letter'' in an alphabet \cite{GSo3}. The field configuration represented by the function $g(\mathbf{x})$ (we will consider only scalar functions here)  is the ``message,'' composed of many field modes. (In principle, there would be infinitely many modes in the continuum, although in a discrete lattice there is always a level of coarse-graining due to UV cutoffs.) The modal fraction gives the relative probability for the occurrence of a specific mode $k$, and the CE measures the information encoded in a given configuration taking into account all modes: in general, the more modes contribute to the Fourier transform of the function $g(\mathbf{x})$, the higher its CE. We can thus associate the CE with a measure of the function's spatial complexity: a single Fourier mode having the lowest CE and hence lowest spatial complexity, while configurations where modes contribute equally maximize CE and thus spatial complexity.

We now search for a relationship between the configuration entropy of excited states of the hydrogen atom and their lifetimes.

\section{Configurational Entropy of Hydrogen Atom and Spontaneous Emission Rates}
\label{CEHatom}
\subsection{Motivation}
\noindent The hydrogen atom is one of the few bound physical systems which were studied analytically since the early development of quantum mechanics. Due to the linearity of the Schr\"odinger equation, the hydrogen wave function can be separated into a radial and an angular part \cite{Sakurai},
\begin{equation}
\Phi_{n\ell m}(r,\theta,\phi) = R_{n\ell}(r)Y_{\ell m}(\theta,\phi),
\end{equation}
with $n,\ell,m$ denoting the principal, angular, and magnetic quantum number, respectively.

\indent As is well-known, an excited state may decay to a lower excited state and eventually to the ground state by emitting a photon of  wavelength $h\nu = \Delta E$, where $\Delta E$ is the energy difference between the two states. In old quantum theory, spontaneous emission is explained phenomenologically using detailed balance of the electron and the radiation field. Einstein \cite{Einstein} introduced the coefficient $A_{if}$, the probability per unit time for the electron to decay from $i\to f$, in terms of $B_{if}$, the coefficient of stimulated emission probability per unit time as,
\begin{equation}
A_{if}\propto \nu^3 B_{if}.
\end{equation}
\indent Spontaneous emission was not thoroughly described until the advent of QED, when the coupling between the electromagnetic field in empty space and the atom was included. Ref. \cite{Weisskopf} gives a purely quantum treatment for spontaneous emission, leading to the same result as that obtained using the Einstein coefficient $A_{if}$, where the initial and final states should obey the selection rules for dipole transitions, $\Delta \ell = \pm 1$ and $\Delta m = 0, \pm 1$. The expression of the transition coefficient $A_{if}$ averaged over angular momentum is,
\begin{equation}
A_{if} = \frac{4\omega^3}{3\hbar c^3}\left (\frac{\hbar^2}{me}\right )^2 |\langle f|r|i \rangle |^2,
\end{equation}
\noindent
where $\omega = 2\pi\nu_{if}$ is the angular frequency for the transition $i\rightarrow f$, $\hbar$ is Planck's constant, $m$ is the electron mass, $e$ its charge, and $\langle f|r|i \rangle$ is the dimensionless transition matrix element averaged over angular momentum.

\indent In what follows, we will relate the stability (lifetime) of the excited states of the hydrogen atom against decay via spontaneous emission to their respective configurational entropy (CE). To do this, we compute the CE of the probability density of various hydrogen-atom excited states and compare the results to the corresponding spontaneous emission rates.   

\subsection{Fourier Transform of the Hydrogen Atom Density Function}
\noindent In order to apply the concept of CE to the hydrogen atom, we first need to compute the Fourier transform of the wave function or of the probability density function of various excited states. The detailed derivation and an example are provided in the Appendix. We repeat the main results here. 

\indent The Fourier transform of the atomic wave function, written in spherical coordinates as $\Phi_{n\ell m}(r,\theta,\phi) = R(r)_{n\ell}Y_{\ell m}(\theta,\phi)$, is:
\begin{equation}
\label{coreresult}
\tilde{\Phi}(k,\alpha,\beta)\propto \sqrt{\frac{(2\ell+1)(\ell-m)!}{(\ell+m)!}}P_{\ell}^{m}\left(\cos\alpha\right)\int_0^{\infty}R_{n\ell}(r)r^2(-kr)^{-\frac{1}{2}}J_{\ell+\frac{1}{2}}\left(-kr\right)dr.
\end{equation}
We could also use spherical Bessel functions in the integrand to write, $(-kr)^{-\frac{1}{2}}J_{\ell+\frac{1}{2}}\left(-kr\right) = j_{\ell}(-kr)$.

\indent To compute the Fourier transform of the hydrogen atom wave function, one would use the specific form of the radial function $R_{n\ell}(r)$ corresponding to an excited state labeled by $\{n,\ell,m\}$ into Eq. \ref{coreresult}.

\indent However, with the interpretation of the probability density as giving spatial information about the electron's position, we find it a more natural quantity to use in the computation of the CE instead of the wave function. This also avoids issues related with the physical interpretation of the wave function. Since the probability density is defined as $|\Phi_{n\ell m}(r,\theta,\phi)|^2$, it has azimuthal symmetry and can be expanded as,
$$|\Phi_{n\ell m}(r,\theta,\phi)|^2 = |R_{n\ell}|^2|Y_{\ell m}|^2 =|R_{n\ell}|^2\sum_{\ell'=0}^{2\ell}A_{\ell'}P_{\ell^{'}}(\cos\theta),$$

\noindent 
where in the last step we used that $|P_\ell^{m}(x)|^2$ is a polynomial in $x$ of order $2\ell$ with even terms only and thus can be expanded in terms of Legendre polynomials of even order from $0$ to $2\ell$. This property allows us to use Eq. \ref{coreresult} to compute the Fourier transform of the probability density function with the following steps: First, expand the angular part of the density function as $|Y_{\ell m}|^2 = \sum_{\ell'=0}^{2\ell}A_{\ell'}P_{\ell^{'}}$ as above. Then, compute the Fourier transform $F_{\ell^{'}}({\bf k})$ of each of the functions $|R_{n\ell}|^{2}Y_{\ell^{'} m=0}$ using Eq. \ref{coreresult} for all $\ell' = 0,...,2\ell$. Because the Fourier transform is a linear functional, sum over all $F_{\ell^{'}}({\bf k})$ with the weight $A_{\ell'}$ computed in the first step and the result is the Fourier transform of the probability density function corresponding to a given state $\{n,\ell,m\}$.

\indent Once we have the Fourier transform of the probability density function, we can calculate the CE as defined in Section \ref{CEsec}.

\subsection{Configurational Entropy of the Hydrogen Atom Excited States}
\noindent We calculate the CE for all states of the hydrogen atom with principal quantum number $n\leq 20$ using the probability density function. As usual, the states are denoted as
$$ 1s = |1,0,0\rangle ; 2s = |2,0,0\rangle ; 2p_{-1} = |2,1,-1\rangle; 2p_0 = |2,1,0\rangle; 2p_{+1} = |2,1,1\rangle...$$
all the way to $n,\ell,m = 20,19,+19$. Each state will have a CE which we write as ${\rm CE}[n,\ell,m]$.

\indent To compute ${\rm CE}[n]$, the CE for each principal quantum number $n$, we compute the individual ${\rm CE}[n,\ell,m]$ for all $n$-degenerate states, sum them,  and average over the degeneracy $n^2$. We refer to this quantity as the ``state-averaged CE,''
\begin{equation}
\label{nCE}
{\rm CE}[n] = \sum_{\ell,m}{\rm CE}[n,\ell,m]/n^2.
\end{equation} 

\noindent For example, ${\rm CE}[n=3]$ is, 
\begin{multline*}
\text{CE}[n=3] = \{\text{CE}[3,0,0]+\text{CE}[3,1,-1]+\text{CE}[3,1,0]+\text{CE}[3,1,1]+ \\
\text{CE}[3,2,-2]+\text{CE}[3,2,-1]+\text{CE}[3,2,0]+\text{CE}[3,2,1]+\text{CE}[3,2,2]\}/9.
\end{multline*}
\noindent The results for $2\leq n\leq 20$ are listed in Table I. 

For the transition probability of spontaneous emission, we use the database of Ref. \cite{db}, which also lists it as a function of principal quantum number $n$, $p(n) = $1/lifetime. This database contains all hydrogen dipole transition probabilities (in units of $10^{8}s^{-1}$) averaged over angular momentum up to $n=200$. As an example, to get the {\it total} transition probability of the $n=2$ state, they sum over the possible dipole transitions
$$p(2p_{+1}\to1s)+p(2p_{0}\to1s)+p(2p_{-1}\to1s),$$
and divide by $n^2 = 4$, which is the degeneracy of the $n=2$ state. This justifies our averaging scheme for the CE of the state $n$ above.

\indent We take one further step and average the transition probability over the number of possible transition routes in terms of the principal quantum number. For instance, the state $n=2$ can only decay to $n=1$ (via two-photon emission in the case of $\ell=0$, but via dipole for $\ell=\pm 1$), while $n=3$ can decay to $n=1$ or $n=2$. We thus define the decay-route averaged transition probability as 
\begin{equation}
\tilde p(n)\equiv p(n)/(n-1).
\label{averaged transition probability}
\end{equation}

\begin{figure}
\includegraphics[width=0.75\linewidth]{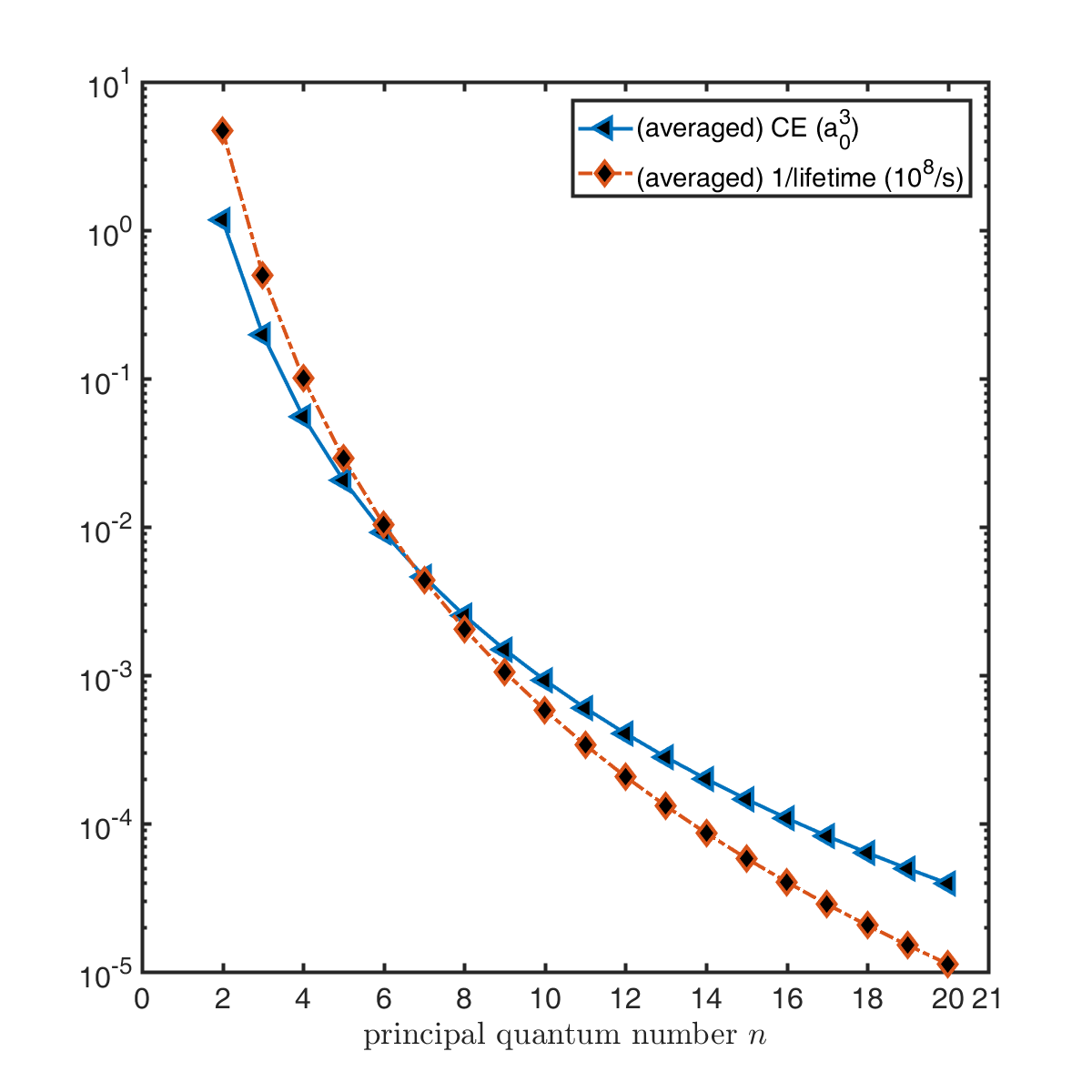}
\caption{Logarithmic plot of averaged transition probability $\tilde p(n)$ (dash-dot line) and state-averaged CE (continuous line) with respect to principal quantum number $n$.}
\label{pCEvsn}
\end{figure}

In Table I we list $\tilde p(n)$ computed from the database of Ref. \cite{db} and the related state-averaged CE for the $n=2,...,20$ states. For clarity, in Fig. \ref{pCEvsn} we plot the numbers from the table, that is, both ${\rm CE}[n]$, the state-averaged configurational entropy computed from the hydrogen probability density
function (continuous line), and the decay-route averaged transition probability $\tilde p(n)$ (dash-dotted line) as functions of principal quantum number $n$.
\begin{table}
 \label{Table1}
\begin{tabular}{|c|c|c|c|c|c|c|c|c|c|c|c|}
\hline
$n$&2&3&4&5&6&7&8&9&10\\
\hline
 ${\tilde p}$ & 4.6967 & 0.4990 & 0.1006 & 0.0289 & 0.0104 & 0.0044 & 0.0020 & 0.0010 & 0.0006 \\ 
\hline
{\rm CE}& 1.1737 & 0.1974 &0.0553 &0.0206 &0.0097 &0.0049 &0.0026 &0.0015 &0.0009 \\
 \hline
\end{tabular}

\vspace{15pt}

\begin{tabular}{|c|c|c|c|c|c|c|c|c|c|c|c|c|}
\hline
$n$&11&12&13&14&15&16&17&18&19&20\\
\hline
 ${\tilde p (10^{-5}) }$ &33.89& 20.69&	13.13&	8.619&	5.821&	4.031&	2.854&	2.060&	1.514&	1.129 \\ 
\hline
{\rm CE $(10^{-5}$)} &60.02 &	40.43	&28.02	&19.98&	14.58&	10.87	&8.240	&6.350	&4.968	&3.942 \\
 \hline
\end{tabular}
\caption{Averaged transition probability ${\tilde p}(n)$ ($10^8/{\rm s}$) defined in Eq. \ref{averaged transition probability} with numbers from the database of Ref. \cite{db} and state-averaged CE for states $2\leq n\leq 20$.}
\end{table}

\begin{figure}
\includegraphics[width=0.75\linewidth]{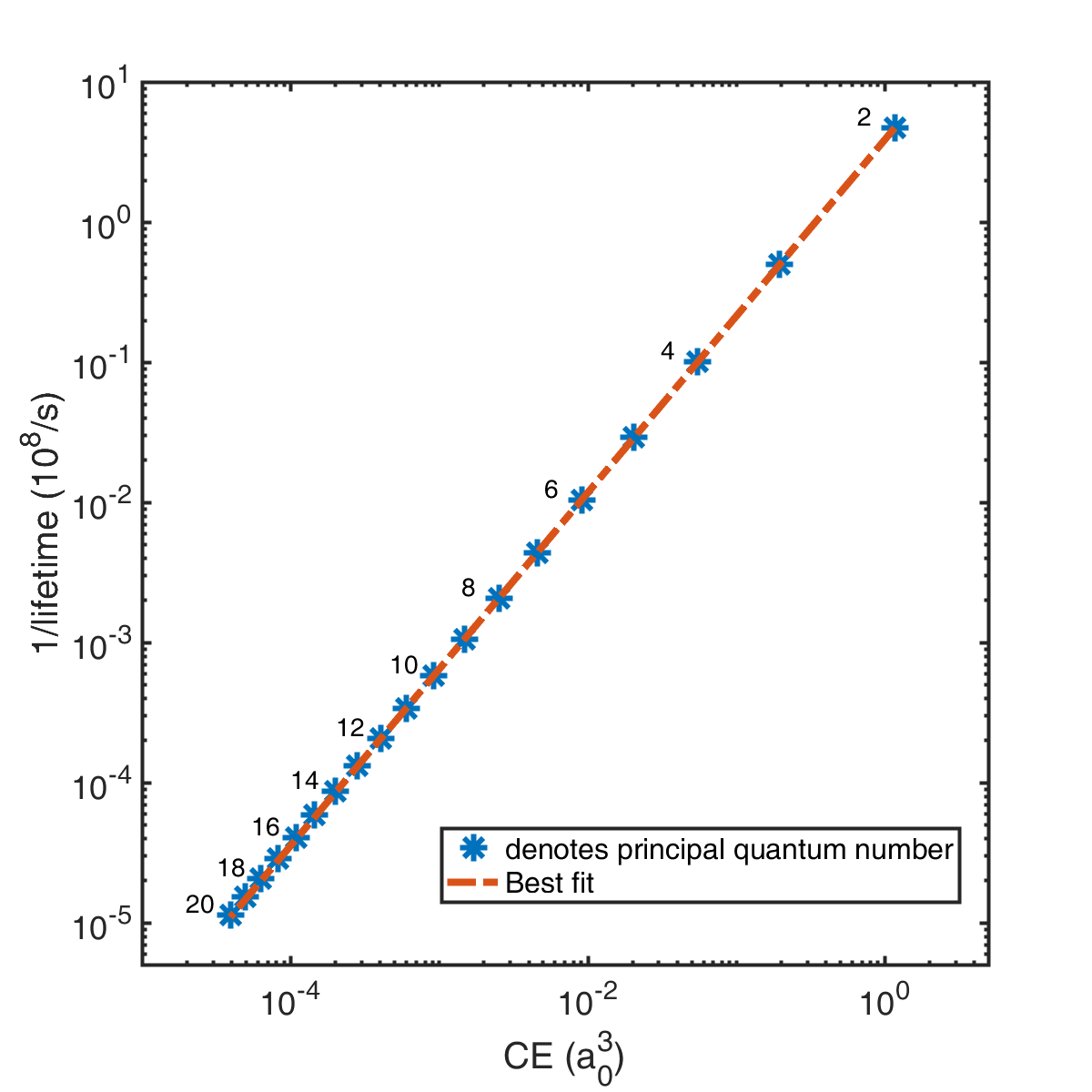}
\caption{Transition probability $\tilde p$ vs. CE. The principal quantum numbers are indicated explicitly. The log-log fit shows the accurate linear relationship between $\tilde p$ and CE. The best fit slope as defined in Eq. \ref{scaling} is $b=1.2595\pm 0.0005$.
}
\label{CE-LFT}
\end{figure}

The monotonic downward trend suggests a possible scaling relation between $\tilde p(n)$ and CE. Indeed, Fig. \ref{CE-LFT} is the log-log plot of $\tilde p(n)$ and the state-averaged CE($n$) with respect to principal quantum number $n$. There is a clear linear fit, which we can write as a scaling relation

\begin{equation}
\tilde p(\text{CE}) = a(\text{CE})^b,
\label{scaling}
\end{equation}
\noindent
where $a$ and $b$ are constants. From a best fit analysis we obtain the exponent $b=1.2595\pm 0.0005$ or $b\simeq 5/4$.

From both Figure \ref{CE-LFT} and Table I we see that the state-averaged configurational entropy ${\rm CE}[n]$ tracks closely the total decay rate from dipole transition probabilities for a principal quantum number $n$. Considering the complexity of the computation of decay rates using analytical techniques, in particular the overlap integrals between initial and final states, we have found that from the perspective of the configurational entropy the information about the overall instability (lifetime) of an excited state with quantum number $n$ is encoded in the averaged sum of the configuration entropy ${\rm CE}[n,\ell,m]$ of the individual degenerate states for that $n$. This is related to the CE of a spatially-bound function being a measure of its localization: the more spatially-localized the function, the higher its CE and the more unstable it is. In the context of atomic physics, this instability is manifested as a shorter lifetime of the state. 

\begin{figure}
\includegraphics[width=0.95\linewidth]{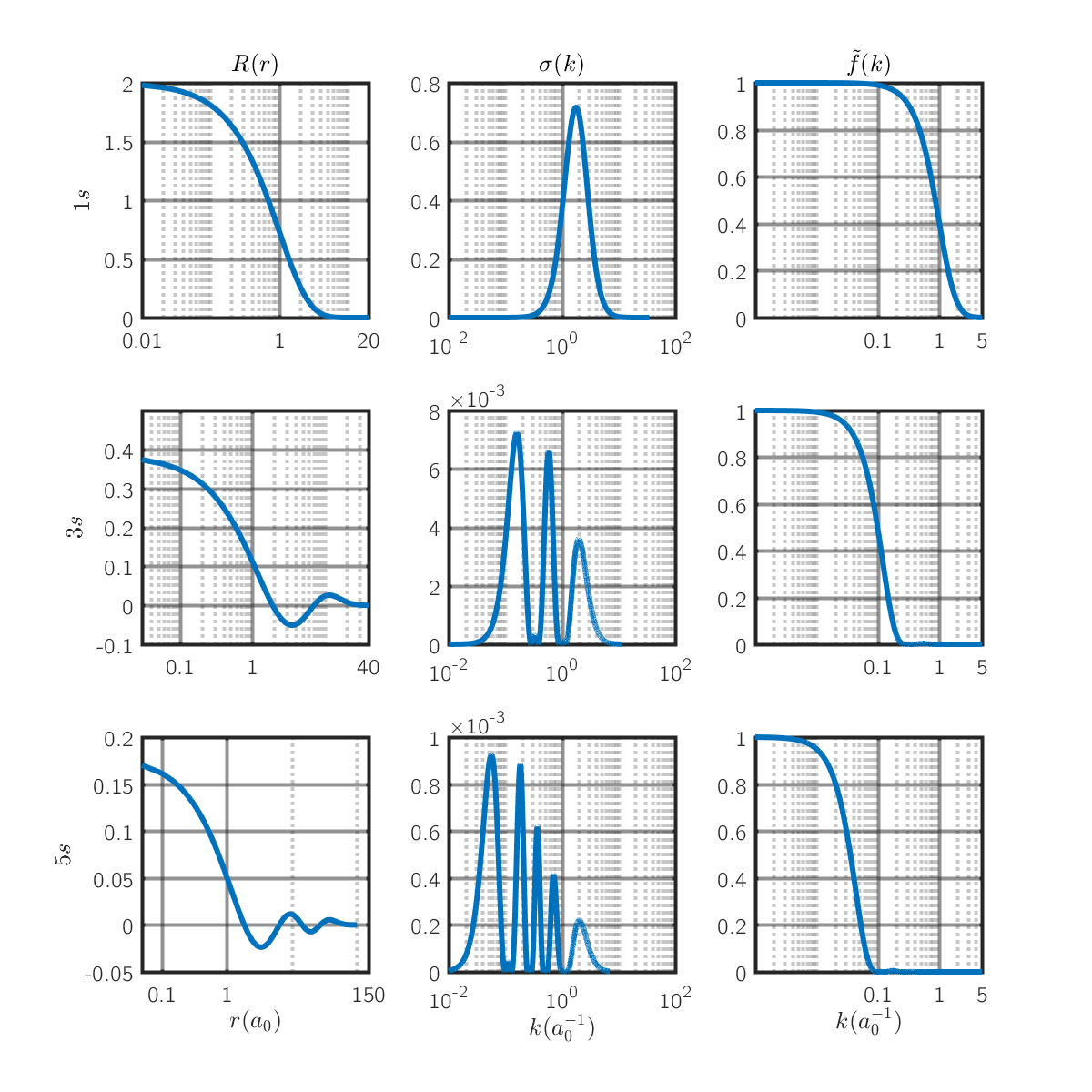}
\caption{From left to right, radial probability density $\Psi_{n00}(r)\Psi_{n00}^*(r)$, CE density $\sigma(k)$, and normalized modal fraction ${\tilde f}(k)$ for  $n=1,~3,~{\rm and}~5$.}
\label{H-atom localization}
\end{figure}

We can see this intuitively by computing the CE of a Gaussian function in $d$ dimensions. As shown in Ref. \cite{GS1}, for a Gaussian function $g(r) = N\exp[-\alpha r^2]$, the normalized modal fraction is $\tilde f(k) = \exp[-k^2/2\alpha]$ (independent of spatial dimensions) and the CE is $S[g] = \frac{d}{2}(2\pi\alpha)^{(d/2)}$. Clearly, as $\alpha\rightarrow \infty$, the Gaussian approaches a delta function in space, and its CE diverges: in that limit, as we know from the Fourier integral of a delta function, all momentum modes would contribute with equal amplitudes. In analogy with thermodynamic entropy, which is maximized at mode equipartition, the configuration entropy is maximized when all modes contribute equally to the modal fraction \cite{GS1}. In information theory, this corresponds to the state of maximal ignorance, a message requiring the largest number of bits to encode. In the context of configurational entropy as applied to one-electron atoms, ``bits'' correspond to the momentum modes of the Fourier transforms of the individual atomic wave functions. Lower values of $n$ imply in higher spatial localization and hence higher CE. This is illustrated explicitly in Fig. \ref{H-atom localization}, where we plot, from left to right, the probability density $\Psi_{n00}(r)\Psi_{n00}^*(r)$, the CE-density $\sigma(k)$, and the normalized modal fraction ${\tilde f}(k)$ for $n=1,~3,~{\rm and}~5$, respectively. As $n$ increases, the states are less localized, and this is reflected in a smaller range of $k$ for ${\tilde f}(k)$ and lower amplitudes for $\sigma(k)$. Our central result is the scaling law relating the $n$-averaged probability density's configurational entropy of a given state with principal quantum number $n$ and its lifetime.
For individual states, the lifetimes don't follow a simple trend, although we observed that it's still possible to extract useful correlations between lifetime and CE for most cases, as long as the lifetimes are sufficiently distinct. For example, for the transitions $2p\rightarrow 1s$ and $3s\rightarrow 2p$ the lifetimes are $1.6\times 10^{-9}$s and $1.6\times 10^{-7}$s, respectively, and their CE's are (in units of $a_0^3$) CE$(2p) = 3.84$ and CE$(3s) = 0.11$, or, averaging over $2\ell+1$ states, CE$(2p) = 1.28$ and CE$(3s) = 0.11$, consistent with a higher CE predicting a more unstable state. For another example, the transitions $4f\rightarrow 3d$ and $3d\rightarrow 2s$ have lifetimes are $7.3\times 10^{-8}$s and $1.56\times 10^{-8}$s, respectively, and their CE's are CE$(4f) = 0.45$ and CE$(3d) = 1.10$, or, averaging over $2\ell+1$ states, CE$(4f) = 0.06$ and CE$(3d) = 0.22$, respectively. Of course, only a more detailed study can determine the efficacy and limitations of these correlations between lifetime and CE for individual states. Still, once we average over the state's degeneracy, we have shown that the trend is apparent. 

\section{Concluding Remarks}
\label{Concluding}
In this paper we examined spontaneous decay of simple one-electron atoms using an information-theoretic approach. Our results rely on an adaptation of 
Shannon's information entropy to spatially-bound mathematical functions, exploring the distribution of momentum modes under the function's Fourier transform. The essential quantity is the configurational entropy of a given quantum state, ${\rm CE}[n,\ell,m]$, which we obtain by first computing its Fourier transform and then extracting its relative distribution of different momentum modes. To obtain the total CE for a given $n$, we summed over all degenerate states and divided by the degeneracy $n^2$. We then compared the CE for that $n$ with the total decay rate via dipole emission for that same $n$ using well-known tabulated results. Our approach allowed us to obtain a scaling law relating the configurational entropy for a given $n$ and the state's total lifetime. We verified that this law holds with accuracy better than 3.7\% within our numerical accuracy for at least $n\leq 20$, showing no obvious increasing deviation for larger $n$. We thus uncovered a novel way of estimating the average lifetimes of one-electron atoms which can, in principle, be applied to any value of $n$. The scaling law presented here only holds for the total state-averaged decay rate and not for individual states, which have varying lifetimes. A more detailed study is needed for individual cases. Still, they illustrate the use of configurational entropy to study the stability of quantum systems. 

For future work, it would be interesting to compute results for even higher $n$s to see if the scaling law holds. Also, a natural extension of the present work is to compute similar relations not only to multi-electron atoms (using, e.g., the Hartree approximation to obtain a workable wave function to compute the relevant Fourier transforms) but also to other quantum systems, from the simple harmonic oscillator to Bose-Einstein condensates. Work along these lines is currently in progress.

\acknowledgments 
MG and NJ are partially supported by a US Department of Energy grant DE-SC001038.\\

\section{Appendix: Fourier Transforms and Numerical Methods}
\label{Appendix}
\subsection{Computation of Fourier Transforms}
\indent In what follows, we describe the derivation of the Fourier transform of an arbitrary function written in spherical coordinates as $\Phi_{n\ell m}(r,\theta,\phi) = CY_{\ell m}(\theta,\phi)R(r)$. Our derivation reproduces results in Ref. \cite{podolsky}. In Ref. \cite{podolsky}, the derivation is dedicated to atomic wave functions with radial part $R_{n\ell}(r)$ with special form $Ce^{-\gamma r}r^{\ell}L_{n+\ell}^{2\ell+1}(2\gamma r)$, where $L_{n+\ell}^{2\ell+1}$ is the generalized Laguerre polynomial. Here, we obtain the Fourier transform for a general radial function $R_{n\ell}(r)$ and carry out the Fourier transform numerically.

\indent Consider general functions in spatial and momentum coordinates $\Phi_{n\ell m}(x,y,z)$ and  $\tilde{\Phi}_{n\ell m}(k_x,k_y,k_z)$, respectively, related to each other by a Fourier transform 
\begin{equation}
\label{atomFFT}
 F(\mathbf{k})=\tilde{\Phi}_{nlm}(k_x,k_y,k_z) = \iiint e^{-i\vec{k}\cdot\vec{x}}\Phi_{n\ell m}(x,y,z)dxdydz.
\end{equation}
\indent We start by writing $(x,y,z)$ and $(k_x,k_y,k_z)$ in spherical coordinates,
\begin{align}
x &= r\sin\theta\cos\phi && k_x = k\sin\alpha\cos\beta \notag \\
y &= r\sin\theta\sin\phi && k_y = k\sin\alpha\sin\beta \notag\\
z &= r\cos\theta && k_z = k\cos\alpha,
\end{align}
\indent and expand $\vec{k}\cdot\vec{x}$ in Eq. \ref{atomFFT} as:
\begin{align}
\label{kxdot}
\vec{k}\cdot\vec{x} = & kr(\sin\theta\cos\phi\sin\alpha\cos\beta\notag\\
&+ \sin\theta\sin\phi\sin\alpha\sin\beta\notag\\
&+ \cos\theta\cos\alpha)\notag\\
= & kr\left(\sin\theta\sin\alpha\cos(\phi-\beta)+\cos\theta\cos\alpha\right).
\end{align}
\indent The general atomic wave function can be written as
\begin{equation}
\label{waveFunction}
\Phi_{n\ell m}(r,\theta,\phi) = (A e^{\pm i m\phi})(B P_{\ell}^{m}(\cos\theta))(Ce^{-\gamma r}r^{\ell}L_{n+\ell}^{2\ell+1}(2\gamma r)),
\end{equation}
\noindent where $A,B,C$, and $\gamma$ are constants independent of coordinates. Ref. \cite{podolsky} provides a detailed derivation for this specific form of the radial function.

\indent We are interested only in the general radial form, which is:
\begin{equation}
\Phi_{n\ell m}(r,\theta,\phi) = (A e^{\pm i m\phi})(B P_{\ell}^{m}(\cos\theta))(R_{n\ell}(r)).
\end{equation}
\indent Using Eq. \ref{kxdot}, we get
\begin{align}\label{totalI}
&\tilde{\Phi}_{n\ell m}(k,\alpha,\beta)  = (A e^{\pm i m\phi})(B P_{\ell}^{m}(\cos\theta))(R_{n\ell}(r))\notag\\
&= AB\iiint e^{-ikr\sin\theta\sin\alpha\cos(\phi-\beta)\pm im \phi-ikr\cos\theta\cos\alpha}P_{\ell}^{m}R_{n\ell}(r)r^2\sin\theta drd\theta d\phi\notag\\
&=AB\int_0^{\infty}R_{n\ell}(r)r^2dr\int_0^{\pi}e^{-ikr\cos\theta\cos\alpha}P_{\ell}^{m}(\cos\theta)\sin\theta d\theta\int_0^{2\pi}e^{-ikr\sin\theta\sin\alpha\cos(\phi-\beta)\pm im\phi}d\phi.
\end{align}
\indent Consider first the $\phi$ integral:
\begin{equation}
\label{I1}
I_1 = \int_0^{2\pi} e^{-ikr\sin\theta\sin\alpha\cos(\phi-\beta)\pm i m \phi}d\phi.
\end{equation}
Introducing $\phi-\beta = \omega$,
\begin{align}
I_1 &=\int_\beta^{2\pi-\beta} e^{-ikr\sin\theta\sin\alpha\cos\omega\pm i m(\omega+\beta)}d\omega\notag\\
&= e^{\pm i m\beta}\int_0^{2\pi}e^{-ib\cos\omega\pm im\omega},
\end{align}
where $b=kr\sin\theta\sin\alpha$. The limits of integration can be changed to $(0,2\pi)$ due to the cyclic property of the integrand. Using the integral expression of the Bessel function
\begin{equation}
J_n(x) = \frac{1}{2\pi}\int_0^{2\pi}e^{\frac{in\pi}{2}}e^{in\tau-ix\cos\tau}d\tau,
\end{equation}
\noindent
and since $e^{\frac{in\pi}{2}}=(i)^n$, we can write Eq. \ref{I1} as
\begin{equation}
I_1 = 2\pi(-i)^m e^{\pm im\beta}J_{m}(b).
\end{equation}
\noindent
Eq. \ref{totalI} becomes
\begin{equation}
\label{totalI-2}
\tilde{\Phi}_{n\ell m}(k,\alpha,\beta) =2\pi AB e^{\pm im\beta}(-i)^{m}\int_0^{\infty}R_{n\ell}(r)r^2dr\int_0^{\pi}e^{-ikr\cos\theta\cos\alpha}P_{\ell}^{m}(\cos\theta)J_m(b)\sin\theta d\theta.
\end{equation}
Consider now the integral
\begin{equation}
I_2 = \int_0^{\pi}e^{-ikr\cos\theta\cos\alpha}P_{\ell}^{m}(\cos\theta)J_m(b)\sin\theta d\theta.
\end{equation}
First, use the generating function defined as
\begin{equation}
(1-2tx+t^2)^{-\nu}\equiv\sum_{\ell=0}^{\infty}C_{\ell}^{\nu}(x)t^{\ell},
\end{equation}
to write the generating function of the Legendre polynomials as
\begin{equation}
\sum_{\ell=0}^{\infty}P_{\ell}(x)t^{\ell} = \frac{1}{\sqrt{1-2tx+t^2}},
\end{equation}
where $P_{\ell}(x)=C_{\ell}^{1/2}$ is the coefficient for $\nu=1/2$.

Apply the operator $(-1)^m(1-x^2)^{\frac{m}{2}}\frac{d^m}{dx^m}$ to both sides and using the definition of the associate Legendre polynomials,
\begin{align}
\sum_{\ell=0}^{\infty}P_{\ell}^{m}(x)t^{\ell}\notag=(-1)^m(1-x^2)^{\frac{m}{2}}(2m-1)!(1-2tx+t^2)^{-\frac{2m+1}{2}}t^m,
\end{align}
whose right-hand side is simply:
\begin{equation*}
(-1)^m(1-x^2)^{\frac{m}{2}}(2m-1)!\sum_{\ell=0}^{\infty}C_{\ell}^{m+\frac{1}{2}}(x)t^{\ell+m}.
\end{equation*}
\indent Equating powers of $t$ we obtain the identity:
\begin{equation}
P_\ell^{m}(x) = (-1)^m(1-x^2)^{\frac{m}{2}}(2m-1)!C_{\ell-m}^{m+\frac{1}{2}}(x).
\end{equation}
Using the identity from Ref. \cite{watson},
\begin{equation}
\int_0^{\pi} e^{iz\cos\theta\cos\alpha}J_{\nu-\frac{1}{2}}(z\sin\theta\sin\alpha)C_{\mu}^{\nu}\left(\cos\theta\right)\sin^{\nu+\frac{1}{2}}\theta d\theta = \left(\frac{2\pi}{z}\right)^{1/2}i^{\mu}\sin^{\nu-\frac{1}{2}}\alpha C_{\mu}^{\nu}(\cos\alpha)J_{\nu+\mu}(z),
\end{equation}
\indent and writing $z=-kr$, $\nu=\frac{1}{2}+m$, $\mu=\ell-m$, we get
\begin{align}
I_2 =& \int_0^{\pi}e^{-ikr\cos\theta\cos\alpha}P_{\ell}^{m}(\cos\theta)J_m(b)\sin\theta d\theta\notag\\
=&(-1)^m(2m-1)!\int_0^{\pi}e^{-ikr\cos\theta\cos\alpha}C_{\ell-m}^{m+\frac{1}{2}}(\cos\theta)J_m(b)\sin^{\frac{m}{2}+1}\theta d\theta\notag\\
=&(2m-1)!(i)^{\ell-m}\left(\frac{2\pi}{-kr}\right)^{\frac{1}{2}}(1-\cos^{2}\alpha)^{\frac{m}{2}} C_{\ell-m}^{m+\frac{1}{2}}\left(\cos\alpha\right)J_{\ell+\frac{1}{2}}\left(-kr\right)\notag\\
=&(-1)^m(i)^{\ell-m}\left(\frac{2\pi}{-kr}\right)^{\frac{1}{2}} P_{\ell}^{m}\left(\cos\alpha\right)J_{\ell+\frac{1}{2}}\left(-kr\right).
\end{align}
\indent The total Fourier transform is:
\begin{equation}
\tilde{\Phi}(k,\alpha,\beta) = AB\int_0^{\infty}R_{n\ell}(r)r^2(-1)^m(i)^{\ell-m}\left(\frac{2\pi}{-kr}\right)^{\frac{1}{2}}P_{\ell}^{m}\left(\cos\alpha\right)J_{\ell+\frac{1}{2}}\left(-kr\right)dr.
\end{equation}
\indent Neglecting imaginary phases and irrelevant constants, and using expressions for $A$ and $B$ from spherical harmonics we obtain: 
\begin{equation}
\label{result}
\tilde{\Phi}(k,\alpha,\beta)\propto \sqrt{\frac{(2\ell+1)(\ell-m)!}{(\ell+m)!}}\int_0^{\infty}R_{n\ell}(r)r^2(-kr)^{-\frac{1}{2}}P_{\ell}^{m}\left(\cos\alpha\right)J_{\ell+\frac{1}{2}}\left(-kr\right)dr;
\end{equation}
or, grouping factors of $-1$,
\begin{equation}
\tilde{\Phi}(k,\alpha,\beta)\propto (-1)^{\ell}\sqrt{\frac{(2\ell+1)(\ell-m)!}{(\ell+m)!}}\int_0^{\infty}R_{n\ell}(r)r^2(kr)^{-\frac{1}{2}}P_{\ell}^{m}\left(\cos\alpha\right)J_{\ell+\frac{1}{2}}\left(kr\right)dr.
\end{equation}
\indent As we noted in Section III.B, this expression can also be written in terms of the spherical Bessel function
\begin{equation}
\tilde{\Phi}(k,\alpha,\beta)\propto \sqrt{\frac{(2\ell+1)(\ell-m)!}{(\ell+m)!}}\int_0^{\infty}R_{n\ell}(r)r^2 P_{\ell}^{m}\left(\cos\alpha\right)j_{\ell}\left(-kr\right)dr.
\end{equation}
\indent Let us look at a simple example where $\ell=1$ and $m=0$. Eq. \ref{result} becomes:
\begin{align}
\tilde{\Phi}(k,\alpha,\beta)\propto& \cos\alpha\int_0^{\infty}R_{n1}(r)r^2(-kr)^{-\frac{1}{2}} J_{\frac{3}{2}}\left(-kr\right)dr\notag\\
\propto&\cos\alpha\int_0^{\infty}R_{n1}(r)r^2\left(\frac{\sin(kr)}{(kr)^2}-\frac{\cos(kr)}{kr}\right)dr\notag\\
\equiv &  F(k)\cos\alpha,
\end{align}
which introduces a $\cos\alpha$ factor in the Fourier transform, leading to the same symmetry in coordinate and momentum space. The modal fraction is then,
\begin{align}
\tilde{f}(k,\alpha,\beta) =& \frac{|F(k)\cos\alpha|^2}{\left(|F(k)\cos\alpha|^2\right)_{max}}\notag\\
=& \cos^2\alpha \frac{|F(k)|^2}{\left(|F(k)|^2\right)_{max}}\notag\\
\equiv& \tilde{f}(k,\alpha)=\tilde{f}\cos^2\alpha,
\end{align}
since the maximum mode must be given by $|\cos\alpha|=1$.

\indent The configurational entropy is (integrating over the azimuthal coordinate),
\begin{align}
S =& -2\pi\iint \tilde{f}(k,\alpha)\log\tilde{f}(k,\alpha) k^2\sin\alpha dkd\alpha\notag\\
=& -2\pi\iint \tilde{f}(k)\cos\alpha^2\log\left(\tilde{f}(k)\cos^2\alpha\right)k^2\sin\alpha dkd\alpha\notag\\
=& -2\pi \left(\frac{2}{3}\int_0^{\infty} \tilde{f}(k)\log\left(\tilde{f}(k)\right)k^2dk-\frac{4}{9}\int_0^{\infty}\tilde{f}(k)k^2dk\right).
\end{align}

\subsection{Numerical Procedures}
Numerical procedures for this paper consist of two main parts: computation of the integral defined in Eq. \ref{coreresult} to obtain the Fourier transform for the probability density function, and computation of the integral defined in Eq. \ref{CE} to obtain the CE. There are four important parameters that affect substantially the accuracy of the numerical procedures: the step sizes $\Delta r$ and $\Delta k$, and the limits of integration for $r$ and $k$. The optimal step size and interval of integration for the radial variable $r$ should be tackled first, as the density function of the hydrogen atom can be generated analytically using the generalized Laguerre polynomial in Eq. \ref{waveFunction}. 

Given that the probability density function for a given state $\{n,\ell,m\}$ has $n-\ell-1$ nodes, we search for the zeroes of the probability density function until we find $n-\ell-1$ roots, except the origin. We stop the integration in $r$ when the probability density function drops below $10^{-7}$ after the peak following the last node. Call this value $r_{\infty}$.
We set the number of steps as $N = 2^9$. The step size is then $r_{\infty}/N$, which we verified yields stable results.

We compute the Fourier transform using $\Delta k=0.1/r_{\infty}$. To determine $k_{\infty}$, we first locate the peaks of $\tilde{f}(k)k^2$. We then set $k_{\infty}$ as the value of $k$ when the amplitude of $\tilde{f}(k)k^2$ drops to $1\%$ of its last peak.

Based on these parameters and the trapezoid approximation, the numerical integrations yield stable results with an error in the $k$ integral controlled by $\Delta k^2$, which is smaller than $10^{-4}$ for small $n$ and $10^{-8}$ for larger $n$. 

%

\end{document}